\documentclass[aps,prd,twocolumn,10pt,showpacs]{revtex4-1}

    \usepackage{graphicx}
    \usepackage{amsmath}
    \usepackage{mathtools}
    \usepackage{siunitx}

    \newcommand{\action}{\mathcal{A}}
    \newcommand{\keld}{\gamma}
    \newcommand{\invpf}{\Phi}
    \newcommand{\ES}{E_\mathrm{S}}
    \newcommand{\eps}{\varepsilon}
    \newcommand{\diff}[1]{\!\!\mathrm{d} #1}
    \newcommand{\pathd}{\mathcal{D}}
    \newcommand{\dd}[2][]{\frac{\mathrm{d} #1}{\mathrm{d} #2}}
    \renewcommand{\vec}{\boldsymbol}
    \DeclareMathOperator{\erfi}{erfi}
    \DeclareMathOperator{\artanh}{artanh}

\usepackage{xcolor}

\begin{document}

% Front matter
    \title{Prefactor in the dynamically assisted Sauter-Schwinger effect}

    \author{Christian Schneider}
    \author{Ralf Sch\"utzhold}
    \email{ralf.schuetzhold@uni-due.de}

    \affiliation{Fakult\"at f\"ur Physik, Universit\"at Duisburg-Essen, Lotharstr. 1, 47057 Duisburg, Germany }

    \date{20 October 2016}

\begin{abstract}
The probability of creating an electron-positron pair out of the quantum vacuum by a strong 
electric field can be enhanced tremendously via an additional weaker time-dependent field.
This dynamically assisted Sauter-Schwinger effect has already been studied in several works. 
It has been found that the enhancement mechanism depends on the shape of the weaker field.
For example, a Sauter pulse $1/\cosh(\omega t)^2$ and a Gaussian profile $\exp(-\omega^2 t^2)$
exhibit significant, qualitative differences. 
However, so far most of the analytical studies were focused on the exponent entering the 
pair-creation probability. 
Here, we study the subleading prefactor in front of the exponential using the worldline 
instanton method. 
We find that the main features of the dynamically assisted Sauter-Schwinger effect,
including the dependence on the shape of the weaker field, are basically unaffected 
by the prefactor. 
%
%In a recent work, the dependence of the dynamically assisted
%    Sauter-Schwinger effect on the shape of the fast, weak pulse was
%    investigated and shown to exhibit significant, qualitative differences
%    between, for example, a Sauter pulse $1/\cosh(\omega t)^2$ and a
%    Gaussian $\exp(-\omega^2 t^2)$. Using the worldline instanton method,
%    these results are now extended to include the subleading fluctuation
%    prefactor. 
    To test the validity of the instanton approximation, we compare
    the number of produced pairs to a numerical integration of the full
    Riccati equation.
    \end{abstract}

    %\keywords{}

    \maketitle

\section{Introduction}
    The Sauter-Schwinger effect is a striking phenomenon predicted by Quantum
    Electrodynamics (QED), that describes nonperturbative pair creation
    from the QED vacuum by a strong electric field~\cite{Sauter1931,
    Sauter1932,Heisenberg1936, Schwinger1951}. 
    Intuitively, one can visualize this process as an electron 
    tunnelling from the Dirac sea to the positive continuum.
    So far, direct 
    experimental verification has not been possible, due to the extremely
    high critical field strength 
    $\ES = m^2c^3/(\hbar q) \approx \SI{1.3e18}{V/m}$ 
    (corresponding to an intensity of $\SI{4.6e29}{W/cm^2}$)
    where pair production is
    expected for a uniform, static electric field.

    An extension that can significantly lower this threshold is
    \emph{dynamical assistance}~\cite{Schutzhold2008}, where an additional
    weak, time dependent field with the frequency scale
    $\hbar\omega\ll2mc^2$ is superimposed
    onto a static, or slowly varying field.
    In~\cite{Linder2015} the impact of different pulse shapes on the
    dynamically assisted Sauter-Schwinger mechanism has been compared,
    by calculating the exponent of the pair production rate, neglecting
    the fluctuation prefactor.

    In the following, we will apply the \emph{worldline instanton method}
    \cite{Feynman1950,Affleck1982,Schubert2001,Kim2002,Dunne2005,Dunne2006a,%
    Schubert2012} to calculate the full pair production rate in the
    dynamically assisted Sauter-Schwinger effect for different shapes of
    time dependent pulses. Section~\ref{sec:method} will give a summary of
    the method, which is then used in section~\ref{sec:application} to
    yield both numerical results without any further approximations and
    analytical estimates in certain parameter regions. In
    section~\ref{sec:riccati} we will present the numerical methods used to
    solve the Riccati equation which gives the exact number of produced
    pairs (up to numerical accuracy).

    We will work in $1+1$ spacetime dimensions throughout, a choice that will
    be explained in section~\ref{sec:riccati}.
    
    This is sufficient to represent the fields considered in this work,
    as we only need one spatial dimension (in which the electric field is 
    oriented) and the temporal dimension (to study time dependent fields)
    \footnote{This does however not mean that the results for $(3+1)$-dimensional
    spacetime are identical, we will highlight the differences in 
    Section~\ref{sec:method}.}.

\section{Worldline instanton method}
    \label{sec:method}
    Let us first briefly review the worldline instanton method
    (for a detailed derivation see, e.g.~\cite{Dunne2005,Dunne2006a}),
    and in particular how the prefactor differs in $1+1$ and $3+1$ 
    spacetime dimensions.

    We start out with the vacuum persistence amplitude,
    i.e., the probability amplitude that an initial vacuum state 
    $|0_\mathrm{in}\rangle$ remains vacuum $|0_\mathrm{out}\rangle$,
    which can be
    expressed using the effective action $\Gamma_\mathrm{M}$,
    \begin{equation}
        \left<0_\mathrm{out}|0_\mathrm{in} \right> = e^{i \Gamma_\mathrm{M}}.
    \end{equation}
    We use the subscript ${}_\mathrm{M}$ to indicate Minkowskian quantities,
    in contrast to the Euclidean versions we will mostly be concerned with
    in the following.

    If the effective action were to gain an imaginary part, the absolute
    value of the vacuum persistence would deviate from one, which can be
    interpreted as the probability amplitude for pair production:
    \begin{equation}
        P_{e^+e^-} = 1 - \left|\left<0_\mathrm{out}|0_\mathrm{in} \right>\right|^2
        \approx2 \Im \Gamma_\mathrm{M}.
    \end{equation}

    After analytic continuation, the Euclidean effective action can be
    expressed using the \emph{worldline path integral}
    \begin{multline}
    \label{eq:fullpathintegral}
        \Gamma[A_\mu] = \int_0^\infty \frac{\mathrm{d}T}{T}e^{-m^2 T}
        \int\diff{^2x^{(0)}}
        \int_{\mathrlap{x(T)=x(0)=x^{(0)}}}\quad\pathd x \\
        \times \exp\left[-\int_0^T\diff{\tau}\left(
            \frac{\dot{x}^2}{4}+iqA\cdot \dot{x}
        \right)\right]
    \end{multline}
    over closed loops $x_\mu(\tau)$ in Euclidean spacetime, where $\mu$ only
    takes on the values $1$ and $2$, so $\dot{x}^2 = \dot{x}_1^2 + \dot{x}_2^2$
    and $A\cdot \dot{x} = A_1\dot{x}_1 + A_2\dot{x}_2$. In this case
    $x_1$ denotes the spatial component (e.g. $z$) and $x_2$ imaginary time.

    Note that we are considering \emph{scalar} QED here, i.e. a complex
    scalar field coupled to the electromagnetic potential. Compared to QED this
    lacks the spin degree of freedom, which for the class of fields studied in
    the following can be shown to result in a trivial factor of $2$
    only~\cite{Dunne2005}.
    
    The worldline instanton approach is a semiclassical approximation
    to~\eqref{eq:fullpathintegral}, by evaluating both the path integral
    and the integral over $T$ using the saddle point method.

    For the path integral, we need to find a path $x_\mu(\tau)$ with
    $x_\mu(T)=x_\mu(0)$ that extremizes
    \begin{equation}
        \label{eq:instantonaction}
        \action[x_\mu](T) = \int_0^T\diff{\tau}\left(
            \frac{\dot{x}^2}{4}+iqA\cdot \dot{x}
        \right)
    \end{equation}
    for a given $T$. The Euler-Lagrange equations for this action functional
    give the equations of motion
    \begin{equation}
        \label{eq:instantoneqs}
        \ddot{x}_\mu = i q F_{\mu\nu} \dot{x}_\nu.
    \end{equation}
    A solution to~\eqref{eq:instantoneqs} that satisfies the periodicity
    conditions is called a \emph{world line instanton}, in analogy to the
    instantons in nonrelativistic quantum tunneling~\cite{Coleman1985}.

    The saddle point approximation includes an additional prefactor,
    arising from the fluctuations around the extremum. For the path
    integral, this amounts to the determinant of a second order
    differential operator~\cite{kleinert2009}. Remarkably, this determinant
    can be found using a finite determinant comprised of solutions to
    a certain initial value problem (for a derivation using methods of
    complex analysis, see~\cite{Kirsten2003}).

    Including the fluctuation prefactor, the saddle point approximation
    of the path integral in \eqref{eq:fullpathintegral} is given by
    \begin{multline}
        \int_{\mathrlap{x(T)=x(0)=x^{(0)}}}\qquad
        \pathd x\ \exp\left[-\int_0^T\diff{\tau}\left(
            \frac{\dot{x}^2}{4}+iqA\cdot \dot{x}
        \right)\right] \\
        \approx \frac{e^{i\theta}}{4\pi T}
        \sqrt{\frac{|\det[\eta^{(\nu)}_{\mu,\mathrm{free}}(T)]|}
            {|\det[\eta^{(\nu)}_\mu(T)]|}}
        \exp\left(-\action[x_\mu^\mathrm{cl}](T)\right).
    \end{multline}
    The $\eta_\mu$ are solutions to the fluctuation equations of motion
    and $\theta$ is the Morse index~\cite{morse1934} of the fluctuation
    operator (omitted here for brevity). In contrast to the $(3+1)$-dimensional
    case, the finite determinants are $2\times2$ and we get a factor of
    $(4\pi T)^{-1}$ instead of $(4\pi T)^{-2}$.
    
    Let us now restrict ourselves to time dependent, homogeneous electric
    fields of constant direction. In particular, we use the
    Euclidean four-potential
    \begin{equation}
        iA_1 = \frac{E}{\omega} f(\omega x_2),
    \end{equation}
    leading to the electric field
    \begin{equation}
        \vec{E}=Ef'(i\omega t)\vec{e}_z.
    \end{equation}

    In this case, both the instanton action and the fluctuation determinant
    can be found, up to quadrature, in terms of the function $f$.

    The remaining $T$-integral can be done explicitly using the saddle
    point approximation as well. The imaginary part of the Minkowski
    effective action is then completely determined by the single function
    \begin{equation}
    \label{eq:gDefinition}
        g(\keld) = \frac{4}{\pi}\int_0^{\chi^*}\diff{\chi}
            \sqrt{1-\frac{1}{\keld^2}f(\keld \chi)^2},
    \end{equation}
    where $\keld=m\omega/(qE)$ is the \emph{Keldysh parameter} and the 
    turning point $\chi^*$
    is determined by the implicit equation
    \begin{equation}
    \label{eq:chiImplicit}
        \keld = f(\keld \chi^*).
    \end{equation}

    Using this function, we find our final expression for the imaginary
    part of the effective action, and thus the pair production rate:
    \begin{multline}
        \label{eq:effectiveAction}
        \Im \Gamma_\mathrm{M}[A_\mu] \approx \\
            \frac{\mathcal{L}}{m^{-1}}
            \sqrt{\frac{E}{\ES}}
            \frac{\sqrt{2}}{8\pi\keld \invpf(\keld)}
            \exp\left(- \pi \frac{\ES}{E} g(\keld)\right),
    \end{multline}
    with the function
    \begin{equation}
    \label{eq:invpfDefinition}
        \invpf(\keld) = \sqrt{-\dd[^2]{(\keld^2)^2}
        \left(\gamma^2g(\keld)\right)}.
    \end{equation}
    The length $\mathcal{L}$ is given by the spatial extent of the electric
    field, for example the focal spot of a laser beam.

    To check the validity of the semiclassical approximation, we turn to a
    single Sauter pulse, where an exact treatment is possible. In this
    case (for details see~\cite{Dunne2006a})
    \begin{equation}
    \label{eq:singleSauterG}
        g(\keld) = \frac{2}{1+\sqrt{1+\keld^2}},
    \end{equation}
    \begin{equation}
        \invpf(\keld)=\frac{1}{\sqrt{2}}\left(1+\keld^2\right)^{-3/4},
    \end{equation}
    and thus
    \begin{multline}
        \label{eq:instantonSingleSauter}
        \Im\Gamma_\mathrm{M}\approx \frac{\mathcal{L}}{m^{-1}}\frac{1}{4\pi}
            \sqrt{\frac{E}{\ES}}\frac{\left(1+\keld^2\right)^{\mathrlap{3/4}}}
                {\keld}\\
            \times\exp\left(-\frac{\ES}{E}\frac{2}{1+\sqrt{1+\keld^2}}\right).
    \end{multline}
    \begin{figure}
        \includegraphics[width=\linewidth]{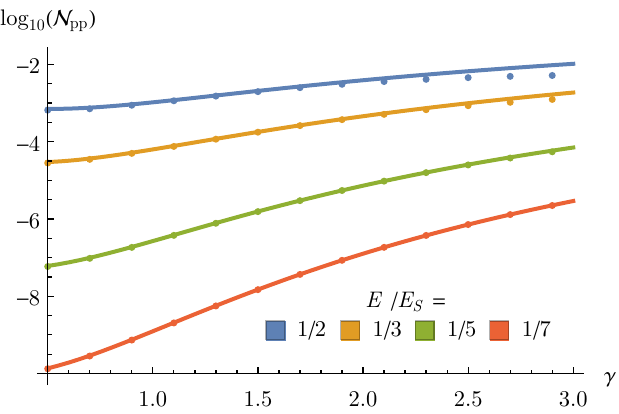}
        \caption{\label{fig:riccatiInstantonSauter}
            Predicted number density of produced pairs by a single Sauter pulse
            with Keldysh parameter $\keld$. The points depict the worldline
            instanton approximation~\eqref{eq:instantonSingleSauter}, the lines
            show the exact analytical result. The field strength increases from bottom
            to top.
        }
    \end{figure}
    In Fig.~\ref{fig:riccatiInstantonSauter} we
    compare~\eqref{eq:instantonSingleSauter} to the known exact solution (see,
    e.g., \cite{Hebenstreit2010}). As expected,
    the semiclassical approximation breaks down for large $E/\ES$, but even for
    $E=\ES/5$ the agreement is excellent, only for $E=\ES/3$ the results start
    to deviate visibly.

    In $3+1$ space-time dimensions, the prefactor differs 
    from~\eqref{eq:effectiveAction} by additional factors~\cite{Dunne2006a}
    \begin{equation}
        \frac{\Im\Gamma_\mathrm{M}^{3+1}}{\Im\Gamma_\mathrm{M}^{1+1}} 
        =
        \frac{V_2}{m^{-2}}\, 
        \frac{E}{\ES}\, 
        \frac{1}{4\pi^2 \dd{(\keld^2)}(\keld^2 g(\keld))}
        \,.
    \end{equation}
    Apart from the obvious two-volume/area $V_2$ and an additional 
    power of $E$, this also includes nontrivial scaling
    with $\keld$. For the single Sauter pulse, this amounts to
    \begin{equation}
        \frac{1}{\dd{(\keld^2)}(\keld^2 g(\keld))} = \sqrt{1+\keld^2},
    \end{equation}
    which does not modify the qualitative behavior.

\section{Dynamically assisted Sauter-Schwinger effect}
    \label{sec:application}
    We now apply \eqref{eq:gDefinition} and \eqref{eq:effectiveAction}
    to the dynamically assisted
    Sauter-Schwinger effect.

    We choose the function
    \begin{equation}
    \label{eq:assistedF}
        f(\chi) = \frac{1}{\rho} \tan(\rho \chi) + \eps h(\chi),
        \quad \eps\ll 1, \quad \rho\ll 1,
    \end{equation}
    representing the sum of a strong, slow field and a weak, faster profile with
    $E_\textrm{weak}/E = \eps$, $\Omega/\omega=\rho$,
    \begin{equation}
        \vec{E} = E(\cosh^{-2}(\Omega t)+\eps h'(i \omega t))\vec{e}_z.
    \end{equation}
    Note that in this case, the \emph{combined} Keldysh parameter
    $\keld=m \omega/(qE)$ compares the frequency scale
    $\omega$ of the weak pulse with  
    the field strength $E$ of the slow, strong field. 
    Note that we do not
    approximate the slow field by a static one (as in~\cite{Linder2015}),
    as the electric field has to vanish for large times for the numerical
    integration of the Riccati equation to work and the limit $\eps\to 0$
    would pose problems in the instanton method.

    In the following, we will choose the following profiles for the fast
    pulses, see also~\cite{Linder2015}
    \begin{itemize}
        \item Cosine $\cos(\omega t)$, $h^\mathrm{cos}(\chi) = \sinh \chi$ 
        \item Gaussian $\exp(-\omega^2 t^2)$, $h^\mathrm{Gauss}(\chi) = \frac{\sqrt{\pi}}{2}\erfi \chi$
        \item Sauter $\cosh^{-2}(\omega t)$, $h^\mathrm{Sauter}(\chi) = \tan \chi$
        \item Lorentzian $(1+\omega^2 t^2)^{-1}$, $h^\mathrm{Lorentz}(\chi) = \artanh \chi$
    \end{itemize}

    First, we will numerically calculate $g(\keld)$ and $\invpf(\gamma)$
    (and thus $\Im\Gamma_\mathrm{M}$) for these pulse shapes $h(\chi)$.
    In section \ref{sec:analytical} we will then present analytical
    approximations for the pair production rate.

    \subsection{Numerical evaluation}
    
    \begin{figure*}[t]
    \centering
    \hfill
    \includegraphics[width=0.4\linewidth]{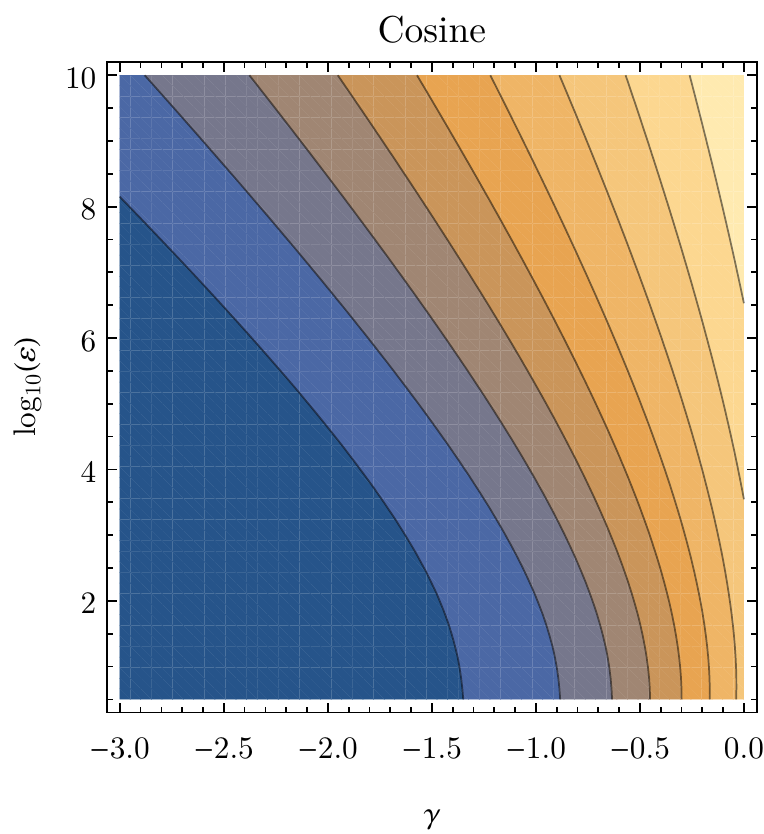}\hfill
    \includegraphics[width=0.4\linewidth]{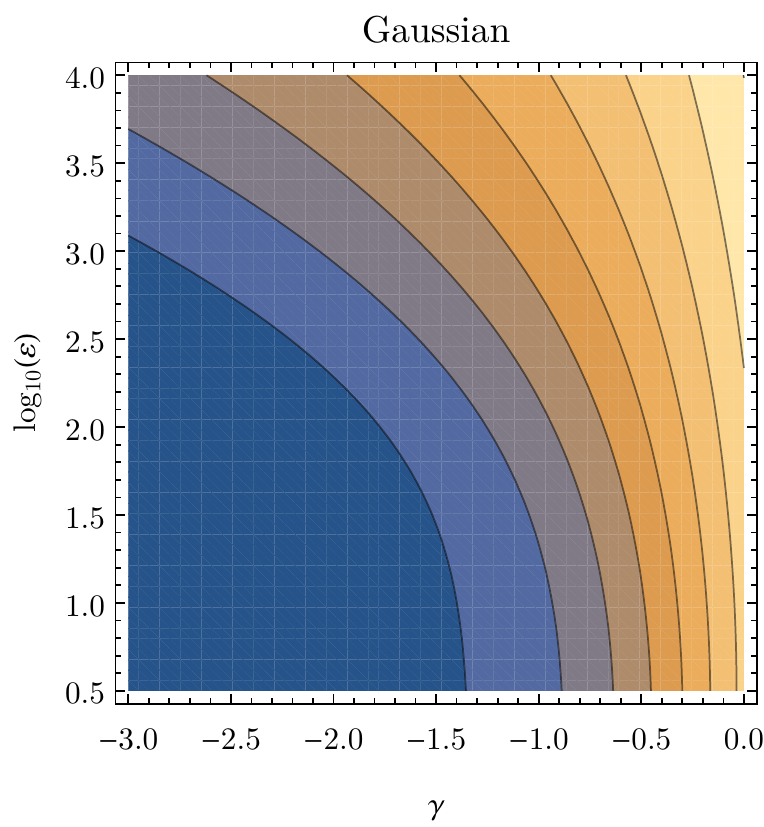}\hfill
    \vspace{5mm}

    \hfill
    \includegraphics[width=0.4\linewidth]{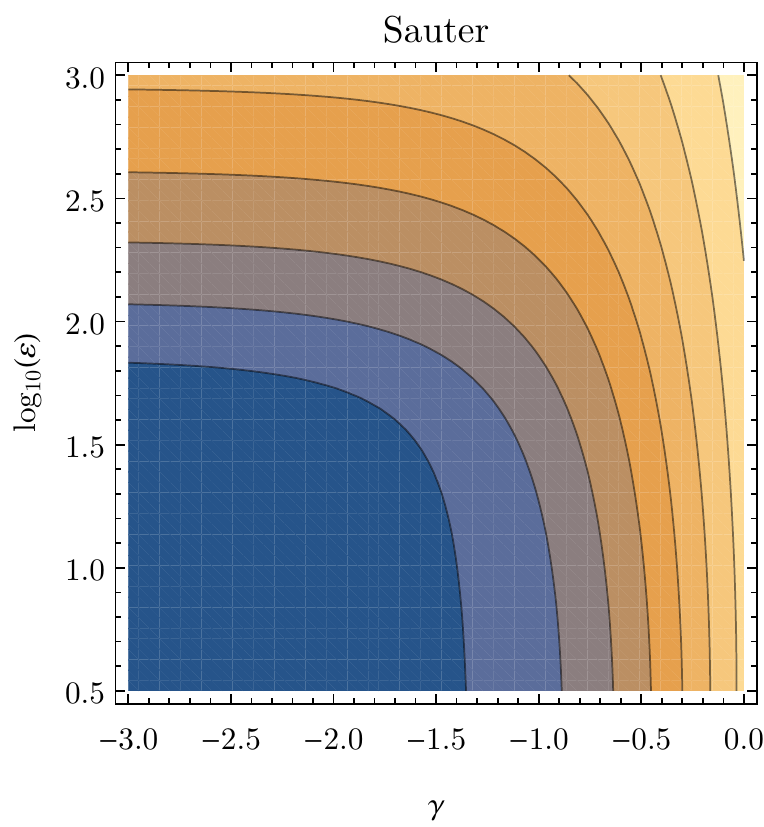}\hfill
    \includegraphics[width=0.4\linewidth]{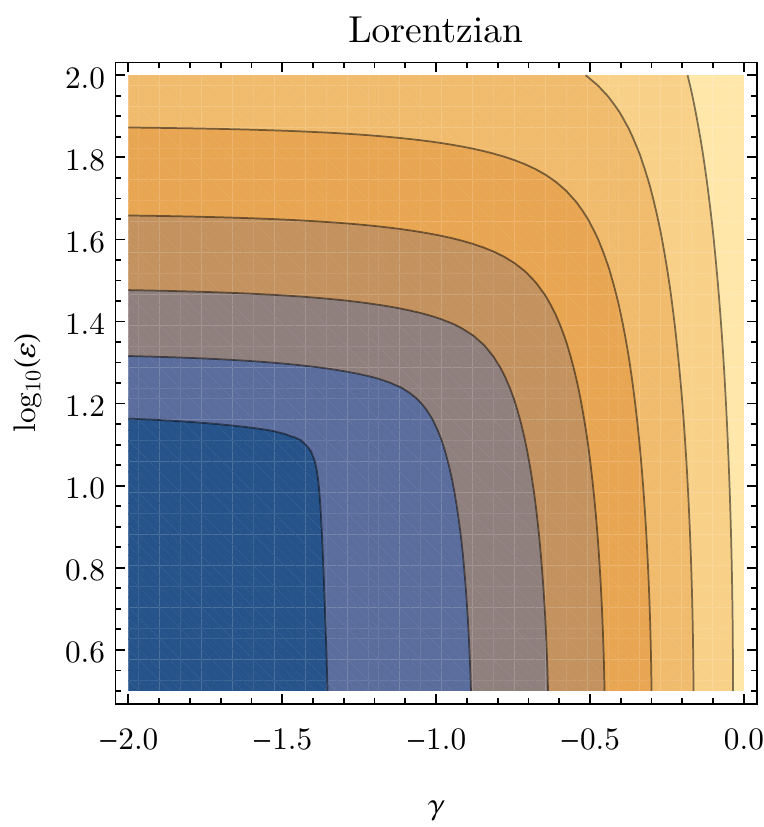}\hfill\mbox{}

    \caption{Imaginary part of the effective
        action~\eqref{eq:effectiveAction}
        in the assisted Sauter Schwinger effect for different pulse shapes
        $h(\chi)$, Keldysh parameters $\keld$ and relative field strengths
        $\eps$. The color scale spans from $10^{-40}$ (blue, bottom left
        corners) to $10^{-16}$ (yellow, top right corners).
        Note the dependence of the threshold $\keld^\textrm{crit}$
        on $\eps$ for the cosine and Gauss profiles, while
        $\keld^\textrm{crit}\approx\text{const.}$ for the Sauter and Lorentz
        profiles.}
    \label{fig:actionContour}
    \end{figure*}

    The effective action~\eqref{eq:effectiveAction} can be evaluated 
    straightforwardly using numerical methods. The only challenge is solving
    the implicit equation~\eqref{eq:chiImplicit}.
    For the cosine and Gaussian profiles, $h$ is smooth and a simple root
    finding algorithm converges using basically any choice of starting point.
    For the other two profiles however, $h$ diverges (at $\chi=\pi/2$ or
    $\chi=1$ respectively), so for small $\eps$, the starting points for the
    root finding method have to be chosen with some care.

    As soon as $\chi^*$ is found, a standard numerical integration routine
    can be used to evaluate~\eqref{eq:gDefinition} and~\eqref{eq:invpfDefinition},
    yielding the effective action~\eqref{eq:effectiveAction}.

    Figure~\ref{fig:actionContour} shows the results of this procedure.
    In all cases there is a region of relatively weak dependence on the
    frequency of the weak field and a region of strong enhancement, as
    soon as the Keldysh parameter crosses a threshold value
    $\keld^\textrm{crit}$. For the cosine and Gauss profiles, this threshold
    depends on $\eps$, while it is approximately constant for the Sauter and
    Lorentz pulses. This is the same behavior as seen in~\cite{Linder2015}
    considering the exponent only, so we can now conclude that the prefactor
    does not change this qualitatively.

\begin{figure}[b]
        \includegraphics[width=\linewidth]{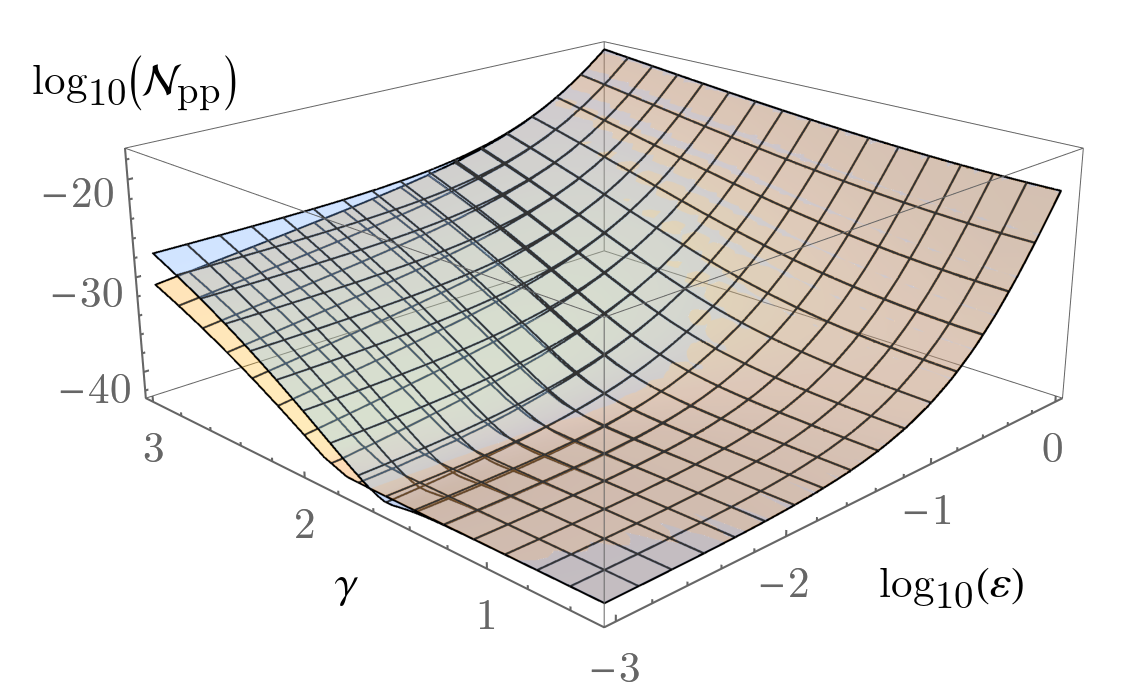}
        \caption{Density of produced pairs from a strong, slow and another
            weak, fast Sauter pulse. The strong field strength is
            $E/\ES = 0.033$, corresponding to an intensity of $\approx\SI{5e26}{W/cm^2}$.
            The blue surface (lying above for large $\keld$ and small
            $\eps$) is the instanton result~\eqref{eq:effectiveAction},
            the orange surface shows the numerical integration of the full
            Riccati equation.
        }
        \label{fig:sautercomp3D}
    \end{figure}

    Now that we have an approximation for the full pair production rate using
    the worldline instanton method, the question that remains is if this
    approximation actually works well for these field configurations. Thus we
    compare it to a solution of the full Riccati equation as outlined in
    section~\ref{sec:riccati}. In Fig.~\ref{fig:sautercomp3D}, we can see
    that both methods agree perfectly below threshold or for large $\eps$. Only
    for small $\eps$ and large $\keld$ the results deviate visibly. This is very
    interesting, because na\"ively one might expect the quality of the
    approximation to depend mainly on the pair production probability (as it
    appears to do in Fig.~\ref{fig:riccatiInstantonSauter}), while in this
    case the interplay between multiple scales leads to a different behavior.

    Furthermore, for small $\eps$ the instanton method predicts a highly
    unintuitive ``dip'' just below the critical value of $\keld$ where the
    pair production actually decreases with increasing Keldysh parameter
    (visible in Fig.~\ref{fig:sautercomp3D} at the lower left edge of the
    displayed surfaces, more obvious in Fig.~\ref{fig:lorentzAna}). The
    numerical solution of the Riccati equation however does not show this
    anomaly, so we can conclude it to be an artifact of the semiclassical
    approximation, again stressing that its validity has to be carefully
    examined in each situation.

    Regardless, the Riccati equation predicts the same qualitative result of
    dynamical assistance above a threshold value of $\keld$, which is roughly
    independent of $\eps$.

    \subsection{Analytical approximations}
    \label{sec:analytical}

    To find analytical expressions for $g(\keld)$ and thus $\invpf(\keld)$
    and $\Im\Gamma_\mathrm{M}$, we can use different approaches for
    $\keld<\keld^\textrm{crit}$ and $\keld>\keld^\textrm{crit}$.

    Below threshold, we can Taylor expand $g(\keld)$ in $\eps$ and $\rho$:
    \begin{align}
        g(\keld) &= \frac{4}{\pi}\int_0^{\chi^*}\diff{\chi}\sqrt{
            1-\left(
                \frac{\tan(\rho\keld\chi)}{\rho\keld}
              + \eps \frac{h(\keld\chi)}{\keld}\right)^2
        } \nonumber\\
        &\approx \frac{4}{\pi}\bigg(
            \int_0^{\chi^*_{\eps=0}}\diff{\chi}
            \sqrt{1-\left(\frac{\tan(\rho\keld\chi)}{\rho\keld}\right)^2}\nonumber\\
          &\qquad\qquad\qquad\qquad- \eps\int_0^1\frac{\diff{\chi}\ \chi}{\sqrt{1-\chi^2}}
          \frac{h(\keld \chi)}{\keld}
        \bigg)\nonumber\\
        &= \frac{2}{1+\sqrt{1+\rho^2\keld^2}} - \frac{4\eps}{\pi\keld}
            \int_0^1\diff{\xi}\ h(\keld\sqrt{1-\xi^2}) \nonumber\\
        &= \frac{2}{1+\sqrt{1+\rho^2\keld^2}} - \frac{4\eps}{\pi\keld} G(\keld).
    \end{align}
    Note that this approximation works only for subcritical $\keld$,
    because otherwise $h(\keld \chi)$ grows large, invalidating the
    expansion. The $\eps$-independent term is just the result for a single
    Sauter pulse with Keldysh parameter $\rho\keld$
    (see~\eqref{eq:singleSauterG}).

    Substituting this expression for $g(\keld)$ in \eqref{eq:invpfDefinition}
    we get
    \begin{multline}
        \invpf(\keld) \approx \\
        \sqrt{
        \frac{\rho^2}{2\left(1+\rho^2\keld^2\right)^{3/2}} +
        \frac{\eps}{\pi\keld^3}\left(\keld^2G''+\keld G' - G\right)}.
    \end{multline}
    Here, it is evident why we kept the slow pulse explicit, instead of
    approximating it as static. If the strong field were independent of time,
    we would have $\invpf\to0$ as $\eps\to0$, leading to
    $\Im\Gamma_\mathrm{M}\to\infty$.
    This is expected, because then for $\eps\to0$, the instanton is not
    confined in the time direction anymore, giving rise to a zero mode. Using
    a slow Sauter pulse for the strong field solves this problem.

    Now all that is left is to calculate $G(\keld)$ for the different pulse
    profiles:
    \begin{description}
        \item[Cosine]
            \begin{equation}
                G^\mathrm{cos}(\keld) = \frac{\pi}{2}I_1(\keld)
            \end{equation}
            \begin{equation}
                \invpf^\mathrm{cos}(\keld) = 
                \sqrt{
            \frac{\rho^2}{2\left(1+\rho^2\keld^2\right)^{3/2}} + \frac{\eps}{2}
                \frac{I_1(\keld)}{\keld}
            }
            \end{equation}
            where $I_\nu$ denotes the modified Bessel functions of the first
            kind.
        \item[Gauss]
           \begin{equation}
                G^\mathrm{Gauss}(\keld) =
                    \frac{\pi\keld}{4} e^{\keld^2/2} \left(
                        I_0(\keld^2/2) - I_1(\keld^2/2)
                    \right) 
            \end{equation}
            \begin{multline}
                \invpf^\mathrm{Gauss}(\keld) = \\ \sqrt{
             \frac{\rho^2}{2\left(1+\rho^2\keld^2\right)^{3/2}} + \frac{\eps}{2}
                e^{\frac{\keld^2}{2}}\left(
                    I_0\left(\frac{\keld^2}{2}\right)
                    + I_1\left(\frac{\keld^2}{2}\right)
                \right)
            }
            \end{multline}
        \item[Lorentz]
           \begin{equation}
                \label{eq:lorentzG}
                G^\mathrm{Lorentz}(\keld) =
                    \frac{\pi}{2} \frac{1-\sqrt{1-\keld^2}}{\keld}
            \end{equation}
            \begin{multline}
                \label{eq:lorentzInvpf}
                \invpf^\mathrm{Lorentz}(\keld) = \sqrt{
            \frac{\rho^2}{2\left(1+\rho^2\keld^2\right)^{3/2}} + \frac{\eps}{2}
                \frac{1}{\left(1-\keld^2\right)^{3/2}}
            }
            \end{multline}
    \end{description}
    For the Sauter profile, it is unfortunately not possible to find a
    closed form expression for the integral in $G(\keld)$, so we can
    not give an analytic expression for its subcritical behavior.

    For $\keld>\keld^\mathrm{crit}$ however, the function $g(\keld)$ can be
    approximated  in the limit $\eps\ll1$ for the Sauter and Lorentzian pulses
    by geometric considerations~\cite{Schutzhold2008,Schneider2015}, leading to
    \begin{equation}
        g^\mathrm{Sauter}(\keld>\pi/2)=
        \frac{2 }{\pi } \arcsin \left(\frac{\pi }{2 \keld }\right) +
        \frac{\sqrt{\keld ^2-\left(\frac{\pi }{2}\right)^2}}{\keld ^2}.
    \end{equation}
    The same method applies to the Lorentzian pulse, the only difference
    being the different value of the critical Keldysh parameter:
    \begin{equation}
        \label{eq:lorentzgAbove}
        g^\mathrm{Lorentz}(\keld>1)=
        \frac{2}{\pi}\left(\arcsin\left(\frac{1}{\keld}\right)+
        \frac{\sqrt{\keld^2-1}}{\keld^2}\right).
    \end{equation}

\begin{figure}
    \includegraphics[width=\linewidth]{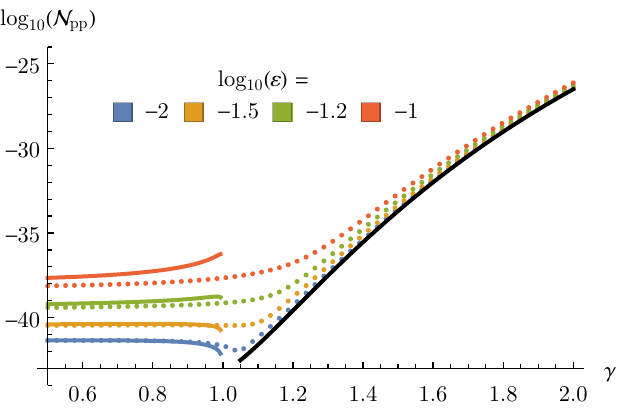}
    \caption{Number density of produced pairs for a Lorentzian
        pulse. The dots represent the numerical results for different
        values of $\eps$, the lines represent the
        approximations~\eqref{eq:lorentzG} and~\eqref{eq:lorentzInvpf}
        for $\keld<1$ and~\eqref{eq:lorentzgAbove} for $\keld>1$. $\eps$
        increases from bottom to top.
    }
    \label{fig:lorentzAna}
    \end{figure}

    Figure~\ref{fig:lorentzAna} shows the numerical calculation of
    $\Im\Gamma^\mathrm{Lorentz}$ for different values of $\eps$,
    the approximations~\eqref{eq:lorentzG} and~\eqref{eq:lorentzInvpf}
    for $\keld<1$ and~\eqref{eq:lorentzgAbove} for $\keld>1$. Far above
    threshold, the pair production rate converges to the
    approximation~\eqref{eq:lorentzgAbove}, independent of $\eps$. Closer
    to the threshold, the geometric approach breaks down for larger
    values of $\eps$, although for $\eps=10^{-2}$ the numerical values
    agree with the approximation very well. As expected, the
    approximation~\eqref{eq:lorentzG} below threshold works better, the
    smaller the expansion parameter $\eps$ gets.

    For $\eps=10^{-2}$ the instanton method again predicts the anomalous
    decrease in pair production at $\keld\approx\keld^\mathrm{crit}$ mentioned
    before.

\section{Numerical solution of the Riccati equation}
    \label{sec:riccati}
    To test the instanton approximations we compared the results to
    a numerical evaluation of the Riccati equation, this chapter explains how
    we obtained these results.
    A brief derivation of the Riccati formalism can be found in~\cite{Linder2015} 
    or in some more detail in~\cite{Dumlu2011c}.
    For a time dependent field pointing in the $z$-direction represented by
    a vector potential $A_3(t)$, we may employ a Fourier transformation 
    in order to account for the spatial dependence of our mode functions. 
    After that, the time-dependence of the instantaneous Bogoliubov coefficients 
    $\alpha_{\vec{k}}$ and $\beta_{\vec{k}}$ is governed by the Riccati equation
    \begin{equation}
    \label{eq:riccati}
    \dot{R}_{\vec{k}} = \Xi_{\vec{k}}(t)\left(
        e^{+2i\phi_{\vec{k}}(t)} + R_{\vec{k}}^2(t) e^{-2i\phi_{\vec{k}}(t)}
    \right)
    \end{equation}
    with $R_{\vec{k}}=\beta_{\vec{k}}/\alpha_{\vec{k}}$ and
    \begin{gather}
    \label{eq:riccatiDefs}
    \begin{gathered}
        \Xi_{\vec{k}}(t) = \frac{q\dot{A}_3(t)\sqrt{m^2+\vec{k}_\perp^2}}
                {2\Omega_{\vec{k}}(t)^2},
            \quad\phi_{\vec{k}}(t) = \int_{-\infty}^{t}\diff{t'}
                \Omega_{\vec{k}}(t'),\\
            \Omega_{\vec{k}}(t) = \sqrt{m^2+\vec{k}_\perp^2+(k_3+qA_3(t))^2},
    \end{gathered}
    \end{gather}
    and the initial condition $R_{\vec{k}}(-\infty)=0$.

    Here, $\vec{k}$ labels the different momentum modes, where $k_3$ denotes
    momentum parallel to the electric field and $\vec{k}_\perp=
    (k_1,k_2)^\top$ the perpendicular momenta. To arrive at the number of
    produced pairs per volume, we need to integrate over all modes:
    \begin{equation}
        \label{eq:riccatiIntegral}
        \mathcal{N}_\mathrm{pp}=\int\frac{\mathrm{d}^3k}{(2\pi)^3}
            \left|R_{\vec{k}}(\infty)\right|^2,
    \end{equation}
    where the factor of $(2\pi)^{-3}$ stems from the choice of normalization
    in the mode decomposition.

    We will, as mentioned in the introduction, work in
    $1+1$ spacetime dimensions, which amounts to setting
    $\vec{k}_\perp=0$. This is due to the fact that we need to find
    $R_{\vec{k}}(\infty)$ for sufficiently many values of $\vec{k}$ to
    approximate the integral in~\eqref{eq:riccatiIntegral}, which is
    computationally intensive. The payoff however is small: While the
    longitudinal momentum spectrum includes important physical effects like
    interference patterns (see, e.g. \cite{Dumlu2011c,Orthaber2011}), the
    perpendicular momenta only amount to a rescaling of the electron mass,
    as is evident in~\eqref{eq:riccatiDefs}. This leads to further exponential
    suppression for $\vec{k}_\perp^2>0$ so they hardly contribute
    to~\eqref{eq:riccatiIntegral} and especially do not modify the qualitative
    response of the pair production rate to the field profile.

    To numerically integrate~\eqref{eq:riccati}, we need to introduce
    dimensionless quantities. First, we choose the vector potential to be
    \begin{equation}
        A_3(t) = \frac{E}{\omega} f(\omega t),
    \end{equation}
    with a dimensionless shape function $f$ and scaling frequency $\omega$.
    We then introduce the quantities
    \begin{equation}
        p=\frac{k_3}{m},\quad\gamma=\frac{m\omega}{qE},\quad
        \tau=\frac{tqE}{m},\quad\mathcal{E}=\frac{E}{\ES}=\frac{qE}{m^2},
    \end{equation}
    leading to the dimensionless Riccati equation, which can be treated
    numerically:
    \begin{subequations}
    \label{eq:riccatiDimless}
    \begin{align}
        \dot{R}_p(\tau)&=
            \frac{\mathcal{E}f'(\gamma \tau)}
            {2\left(1+\left(p+f(\gamma\tau)/\gamma\right)^2
            \right)}\nonumber\\
            &\qquad\times \left(
                e^{i2\varphi_p(\tau)/\mathcal{E}}
                    +R_p^2(\tau)e^{-i2\varphi_p(\tau)/\mathcal{E}}
            \right), \\
        \dot{\varphi}_p(\tau)&=\sqrt{1+
            \left(p+f(\gamma\tau)/\gamma\right)^2}.
    \end{align}
    \end{subequations}
    While $\varphi_p(\tau)$ can of course be obtained immediately in terms of an
    integral, this can usually not be done analytically. Now, since we would
    like to use a variable step width integration algorithm to
    solve~\eqref{eq:riccatiDimless}, we do not know in advance for which values
    of $\tau$ we need $\varphi_p(\tau)$, so we cannot efficiently precompute
    the integral. Thus, it is most convenient to directly solve the system of
    equations~\eqref{eq:riccatiDimless} in lockstep.

    Of course, we cannot perform infinitely many integration steps to find
    $R(\infty)$ from $R(-\infty)=0$, but have to choose a sufficiently long
    time range $\tau\in[-\mathcal{T},\mathcal{T}]$, so that
    $E(|\tau|>\mathcal{T})$ is negligible compared to $E(|\tau|<\mathcal{T})$.
    Then, using the initial condition $R_p(-\mathcal{T})=0$,
    $\varphi_p(-\mathcal{T})=0$ we arrive at the number of produced pairs per
    Compton length
    \begin{equation}
        \mathcal{N}_\mathrm{pp} = \frac{1}{m^{-1}}
            \int\frac{\mathrm{d}p}{2\pi} \left|R_p(\mathcal{T})\right|^2.
    \end{equation}

    Unfortunately, actually obtaining $R_p(\mathcal{T})$ for large $\mathcal{T}$
    and small values of $\mathcal{E}$ is far from trivial. Since the
    exponentials in the Riccati equation oscillate wildly with a frequency of
    order $\mathcal{E}^{-1}$, the step width $\Delta\tau$ has to be sufficiently
    small, which means many steps have to be taken to reach $\mathcal{T}$.
    Using a standard ODE integration algorithm \footnote{Predictor-Corrector
    Adams method, default algorithm of Wolfram Mathematica's \texttt{NDSolve}
    function}, for fields weaker than $\approx \ES/10$ machine precision is
    not sufficient anymore and rounding errors begin to dominate.

    Instead, we use the software package TIDES~\cite{Abad2012} which is based
    on the Taylor series method and supports the GNU MPFR~\cite{Fousse2007}
    library for multiple precision arithmetic, allowing us to integrate
    the Riccati equation using as many significant digits in the calculation
    as needed to give accurate results.

\begin{figure}
        \includegraphics[width=\linewidth]{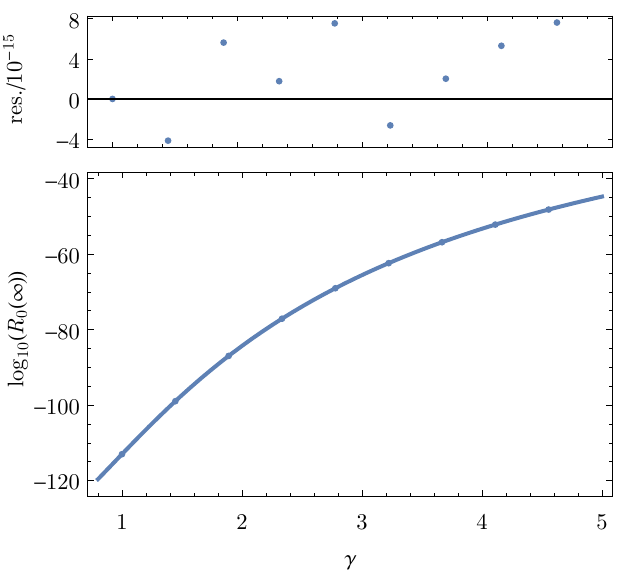}
        \caption{\label{fig:sauterBench}
            Comparison of the analytic solution (line) and numerical results
            (points) of the Riccati equation for a single Sauter pulse with
            $E=\ES/100$ and $k_3=0$. The top plot shows the relative deviation
            from the analytic result.
        }
    \end{figure}

    As a benchmark, we calculated $R_{p=0}(\mathcal{T})$ for a single
    Sauter pulse, which we can again compare to the known analytical result.
    Indeed, using very little computational resources \footnote{About one minute
    per data point on a typical desktop PC}, it is possible to reproduce the
    analytic solution for $p=0$ and various values of $\gamma$ with a relative
    error of less than $10^{-14}$ for a field strength of $E=\ES/100$, see
    Fig.~\ref{fig:sauterBench}.

\section{Summary and conclusion}
    Using the worldline instanton method, we have been able to
    numerically calculate and find analytical approximations
    for the pair production rate in the dynamically assisted
    Sauter-Schwinger effect. Building on~\cite{Linder2015}, this
    now includes the quantum mechanical fluctuation prefactor, which
    is shown not to counteract the mechanism of dynamical assistance.

    Comparing the different pulse shapes considered, they all
    exhibit a similar qualitative behavior. This includes a region
    of negligible dependence on the time dependent field up to
    a threshold value of the Keldysh parameter $\keld$, beyond which
    the pair production rate is exponentially enhanced.

    This threshold $\keld^\textrm{crit}$ is independent of $\eps$
    for the Sauter and Lorentzian pulses, in contrast to the
    sinusoidally varying field and Gaussian pulse which is caused
    by the different analytic structure of the field profiles.

    Furthermore, using a numerical integration of the Riccati equation,
    we have shown that these results are not an artifact of the semiclassical
    approximation, but are present in a full numerical simulation as well.

\bibliography{library}

\end{document}